\documentclass[acmtog,nonacm,screen]{acmart}

\usepackage{booktabs} 

\usepackage[ruled]{algorithm2e} 

\SetAlFnt{\small}
\SetAlCapFnt{\small}
\SetAlCapNameFnt{\small}
\SetAlCapHSkip{0pt}

\acmJournal{TOG}

\setcopyright{none}

\usepackage{comment}

\usepackage{pifont}

\usepackage{multirow}
\usepackage{mwe} 
\usepackage{subcaption} 
\usepackage{graphicx}

\usepackage{bm}

\renewcommand{\vec}[1]{\bm{#1}}
\newcommand{\mat}[1]{\bm{#1}}
\newcommand{\set}[1]{\mathcal{#1}}

\usepackage{enumitem}

\definecolor{best}{rgb}{1, 0.7, 0.7}
\definecolor{second}{rgb}{1, 0.85, 0.7}

\let\originalleft\left
\let\originalright\right
\renewcommand{\left}{\mathopen{}\mathclose\bgroup\originalleft}
\renewcommand{\right}{\aftergroup\egroup\originalright}

\usepackage{hyperref}
\usepackage{cleveref}
\begin{document}
\title{PiG-Avatar: Hierarchical Neural-Field-Guided Gaussian Avatars}

\author{Julian Kaltheuner}
\orcid{0000-0002-5218-1638}
\affiliation{%
 \institution{University of Bonn}
 \country{Germany}}
\email{kaltheun@cs-uni-bonn.de}

\author{Jan Spindler}
\orcid{0009-0003-7852-7654}
\affiliation{%
 \institution{University of Bonn}
 \country{Germany}}
\email{jspindle@cs.uni-bonn.de}

\author{Sina Kitz}
\orcid{0009-0006-3461-0570}
\affiliation{%
 \institution{University of Bonn}
 \country{Germany}}
\email{kitz@cs.uni-bonn.de}

\author{Patrick Stotko}
\orcid{0000-0002-2608-0278}
\affiliation{%
 \institution{University of Bonn}
 \country{Germany}}
\email{stotko@cs.uni-bonn.de}

\author{Reinhard Klein}
\orcid{0000-0002-5505-9347}
\affiliation{%
 \institution{University of Bonn}
 \country{Germany}}
\email{rk@cs.uni-bonn.de}


%
%


\begin{abstract}

Realistic animatable avatars of clothed humans are fundamental to virtual production, telepresence, and immersive experiences, yet reconstructing them from video remains challenging: clothing exhibits complex topology, layered geometry, and non-rigid dynamics that exceed what template-bound representations can faithfully capture. Existing Gaussian avatar methods typically parameterize geometry on a body-template surface, which entangles the avatar's representation space with the template's deformation space. Consequently, the template's topology and sampling density impose a hard constraint on the representable geometry, precluding multi-layered garments, thick off-body clothing, and open-boundary surfaces unless explicit, case-specific mechanisms are introduced.

We present PiG-Avatar, which addresses this limitation by using the parametric body model solely for kinematic transport, while representing the avatar as Gaussians anchored in a volumetric canonical space governed by a continuous neural field. This decouples representation from template topology, avoiding the geometric constraints of surface-based parameterizations. Kinematic coherence is maintained through 3D barycentric anchor transport, which guides motion without constraining geometry and allows anchors to deviate freely from the template surface, yielding dense, stable temporal surface correspondences by construction. To make this unconstrained formulation tractable, we introduce dual-level spatially coherent optimization, combining Sobolev-preconditioned neural-field updates with a novel KNN-based preconditioning of canonical anchor geometry. Together, these mechanisms induce an emergent self-organization of anchor density: anchors migrate toward regions of high curvature, appearance variation, and non-coherent motion without explicit heuristics. As a result, complex clothing geometry and layered surfaces emerge as natural, high-fidelity outputs. This single representation further supports hierarchical reconstruction across multiple levels of detail, with coarse-level supervision propagating to finer levels through the shared field and coupled anchor graph. On established benchmarks featuring subjects with complex clothing and challenging non-rigid motion, PiG-Avatar achieves state-of-the-art rendering quality, generalizes robustly to imperfect body model initialization, and renders in real time across all detail levels.

\end{abstract}

\begin{teaserfigure}
    \centering
    \includegraphics[width=\linewidth, trim=70 67 300 130, clip]{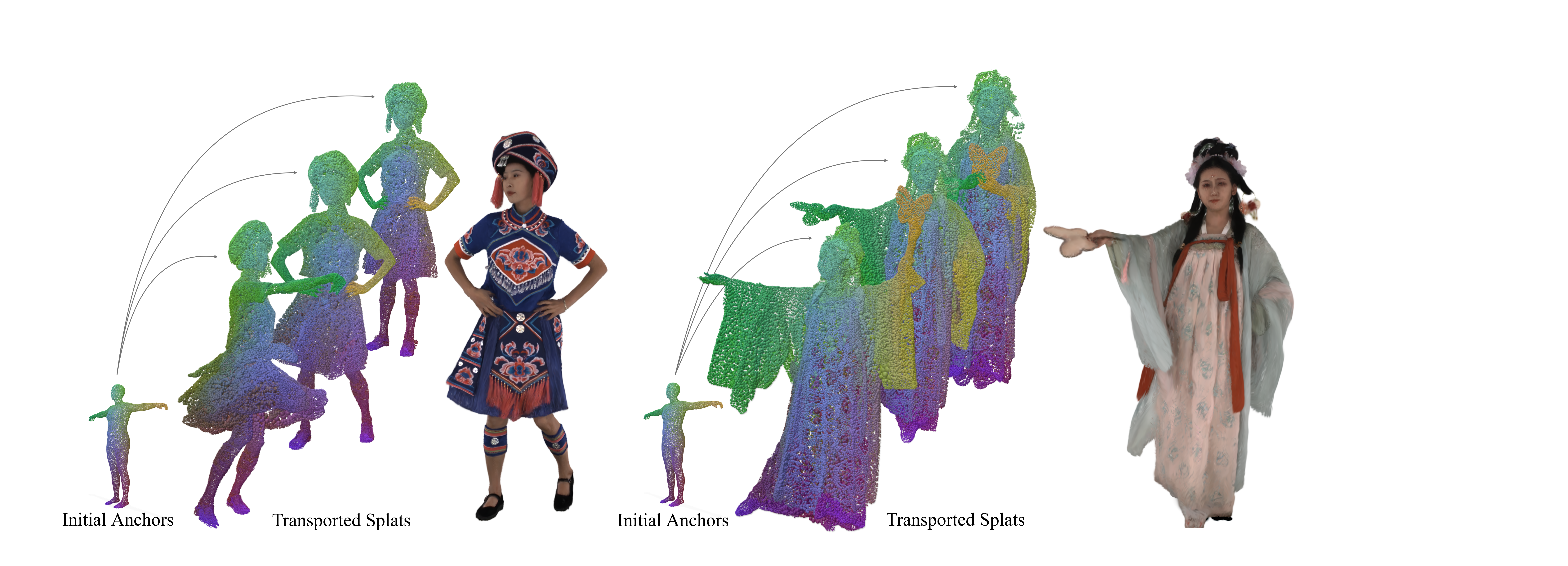} 
    \caption{ We present \emph{PiG-Avatar}, a Gaussian avatar method that decouples representation and deformation: the parametric model provides only kinematic transport, while a canonical neural field over learnable anchors is independent of template topology. Anchors self-organize in canonical space and, combined with time-conditioned neural features, produce temporally consistent posed splats, enabling complex, layered clothing and non-rigid motion. }
    \label{fig:teaser}
\end{teaserfigure}

\maketitle

\section{Introduction}

Realistic animatable human avatars are a key component of virtual production, telepresence, gaming, and immersive AR/VR experiences. A long-standing goal in computer graphics and vision is therefore to reconstruct digital humans from video in a form that is both visually faithful and controllable for animation. While remarkable progress has been made in recent years, modeling clothed humans remains challenging due to the complex geometry and motion of garments. Loose clothing, layered structures, wrinkles, and topology changes often violate the assumptions underlying standard articulated human representations.

Parametric body models such as SMPL~\cite{loper2015smpl}, SMPL-X~\cite{SMPL-X:2019}, and related formulations~\cite{xu2020ghum, ferguson2025mhr} provide compact and highly controllable representations of human pose and shape. Because they establish a consistent articulated structure across time, they have become a common foundation for animatable avatar reconstruction. However, these models represent only minimally clothed bodies and cannot directly capture the geometry of real clothing. Earlier methods therefore augmented the body model with displacement surfaces, textures, or dynamic mesh refinements, but the resulting representations remained closely tied to the topology and resolution of the underlying template. Neural scene representations substantially increased the realism achievable for human reconstruction. Methods based on implicit fields and neural rendering~\cite{park2019deepsdf, mescheder2019occupancy, mildenhall2020nerf} can represent highly detailed geometry and appearance while modeling complex non-rigid motion through learned deformation fields~\cite{Peng_2021_CVPR, Peng_2021_ICCV, liu2021neuralactor, Weng_2022_CVPR, Mu_2023_ICCV, li2023posevocab}. More recently, 3D Gaussian Splatting (3DGS)~\cite{kerbl20233d} has enabled explicit scene representations with real-time rendering performance, making Gaussian-based approaches increasingly attractive for animatable avatars.

Most existing Gaussian avatar methods rely on the body template not only for articulation, but also as the parameterization domain of the representation itself. Gaussian primitives are commonly attached directly to the template mesh or parameterized in UV space defined on the body surface~\cite{Hu_2024_CVPR, jiang2025uv, kwon2024generalizable, Pang_2024_CVPR, Qian_2024_CVPR, Kocabas_2024_CVPR, Hu_2024_CVPR_Gau}. This design provides stable animation and temporal consistency, but it also introduces a strong geometric bias: the structure of the representation inherits the topology, sampling density, and deformation behavior of the body model. Consequently, representable geometry is implicitly restricted to deformations of the template surface itself. In practice, this makes it difficult to faithfully represent garments that deviate substantially from the body surface, such as skirts, coats, layered clothing, or loose fabric with complex motion. Existing approaches often address these limitations through specialized clothing templates, additional deformation modules, or carefully initialized canonical meshes~\cite{li2023animatable, Chen_2025_CVPR, lin2024layga, lee2025mpmavatar}, increasing both modeling complexity and dependence on accurate priors.

In this work, we explore a different formulation. Instead of treating the body template as the geometric domain of the avatar representation, we use it only to provide kinematic guidance for motion. Our method, \emph{PiG-Avatar}, represents the avatar using Gaussian anchors embedded directly in a canonical volumetric space, where geometry and appearance are modeled through a shared continuous neural field. The anchors are initialized with respect to a parametric articulated body model to obtain temporally consistent motion, but they are not constrained to remain on the template surface. As optimization proceeds, the anchors are free to reorganize in 3D space according to the observed geometry and motion of the subject, while the underlying body model parameters are refined to further strengthen motion alignment and stabilize this evolution. This joint refinement supports a more structured organization of the canonical representation without limiting anchors to the template surface, but it also weakens the geometric regularization typically provided by the template. As a result, the canonical representation can become unstable or spatially incoherent, introducing a new optimization challenge: without the structural regularization imposed by a template surface, the representation must instead maintain coherence through learning alone. To address this, we enforce coherence at two complementary levels. First, updates of the shared canonical field are spatially smoothed through Sobolev preconditioning~\cite{kaltheuner2026neu}, encouraging coherent feature evolution in the latent representation. Second, we introduce a neighborhood-aware optimization scheme directly on the canonical anchor graph, coupling nearby anchors during geometric refinement. Together, these mechanisms encourage stable organization of the representation while preserving the flexibility needed to capture complex clothing geometry and non-rigid motion.

An additional advantage of our formulation is that the canonical representation remains temporally consistent by construction. Since anchors maintain stable identities over time while being transported through the articulated motion model, the method naturally produces dense temporal correspondences, as depicted in \Cref{fig:teaser}, without requiring learned deformation tracking or post-processing alignment, directly enabling temporally consistent editing and transfer operations. Furthermore, the same canonical representation can be evaluated at multiple resolutions, allowing hierarchical avatar reconstruction within a single shared framework. We demonstrate that PiG-Avatar reconstructs high-fidelity animatable avatars with challenging clothing and motion while supporting real-time rendering. Experiments on established benchmarks show strong visual quality, robustness to imperfect body model initialization, and consistent performance across multiple levels of detail.

Our main contributions are:
\begin{itemize}

\item We identify the conflation of representation space and deformation space as the fundamental limitation of template-bound Gaussian avatar methods, and present PiG-Avatar, which achieves complete architectural decoupling: the parametric model provides only kinematic transport, while the canonical representation, a volumetric neural field over learnable anchor points, is independent of template topology. As a structural consequence, complex and layered clothing geometry emerges without case-specific provisions.


\item We introduce a \emph{3D barycentric anchor transport} mechanism that preserves kinematic coherence by consistently transporting Gaussian anchors through the articulated motion space, enabling stable and temporally consistent correspondences under deformation.


\item We propose a \emph{dual-level spatially coherent optimization} strategy that builds on Sobolev-preconditioned latent field updates and introduces a second preconditioning level on canonical anchor geometry via a fixed KNN graph. This design enforces spatial coherence simultaneously in the implicit field and explicit anchor positions, making topologically unconstrained canonical Gaussian optimization tractable without geometric regularization terms.

\item We present a single unified hierarchical representation that is learned from a single model and supports progressive level-of-detail reconstruction and real-time rendering.

\end{itemize}

Code will be released upon publication.

\section{Related Work}

\subsection{Implicit Neural Human Avatars}

Neural rendering approaches have substantially advanced the reconstruction of
animatable human avatars from images and videos. Many methods build upon
parametric body models such as SMPL and SMPL-X~\cite{loper2015smpl, SMPL-X:2019},
which provide compact articulated priors for pose and shape and are commonly
used to guide deformation between canonical and posed spaces.

A large body of work represents humans using implicit neural fields based on
Neural Radiance Fields (NeRF). Early approaches such as Animatable NeRF~\cite{Peng_2021_ICCV},
A-NeRF~\cite{NEURIPS2021_65fc9fb4}, and NeuMan~\cite{jiang2022neuman}
learn canonical radiance fields that are animated through inverse skinning or
pose-conditioned deformation fields. Subsequent works improve deformation modeling
through structured local representations~\cite{zheng2022structured},
bidirectional deformation mappings~\cite{yu2023monohuman},
or neural blend weight formulations~\cite{Peng_2021_ICCV}.

Several methods additionally introduce pose-aware latent representations to better model appearance variation. NeuralBody~\cite{Peng_2021_CVPR} associates latent features with SMPL vertices, while NeuralActor~\cite{liu2021neuralactor} learns residual pose-dependent deformations on top of articulated priors. Later approaches, such as ActorsNeRF~\cite{Mu_2023_ICCV}, PoseVocab~\cite{li2023posevocab}, and TexVocab~\cite{Liu_2024_CVPR}, further improve pose-conditioned appearance modeling via structured latent embeddings or texture vocabularies.

Another line of work focuses on canonical implicit surface representations instead of volumetric radiance fields. Methods such as SCANimate~\cite{Saito_2021_CVPR} and SNARF~\cite{Chen_2021_ICCV} improve generalization by learning deformation fields in canonical space. TAVA~\cite{li2022tava} further relaxes reliance on predefined templates by learning canonical volumetric representations directly from observations. Generative approaches, such as AvatarGen~\cite{zhang2022avatargen} and GetAvatar~\cite{zhang2023getavatar}, further explore latent canonical geometry spaces for controllable avatar synthesis. Despite their high visual fidelity, implicit neural avatar methods generally require computationally expensive volumetric rendering and dense sampling during both training and inference.

\subsection{Gaussian-based Human Avatars}

Explicit scene representations based on 3D Gaussian Splatting~\cite{kerbl20233d}
have recently emerged as an efficient alternative to volumetric neural rendering.
By representing scenes as collections of anisotropic Gaussian primitives,
3DGS enables high-quality rendering with substantially faster training and
real-time inference, making it particularly attractive for animatable human avatars.

Most Gaussian avatar methods tightly couple primitives to a parametric human template. Surface-bound approaches such as 3DGS-Avatar~\cite{Qian_2024_CVPR}, HUGS~\cite{Kocabas_2024_CVPR}, GauHuman~\cite{Hu_2024_CVPR_Gau}, and SGIA~\cite{zhao2025surfel} attach Gaussians directly to template surfaces and animate them via linear blend skinning. GART~\cite{Lei_2024_CVPR} learns articulated Gaussian templates initialized on a parametric surface and animated via learnable blend skinning, with limited freedom to deviate from the template geometry. Some methods additionally optimize skinning weights or pose refinements to improve deformation quality~\cite{Lei_2024_CVPR,zhao2025surfel}. Other approaches extend the underlying body representation itself to better capture clothing and secondary motion, for example through clothed templates or learned deformation modules~\cite{Chen_2025_CVPR,lee2025mpmavatar}.

A related direction parameterizes Gaussian attributes in the UV domain of the template surface. GaussianAvatar~\cite{Hu_2024_CVPR}, Animatable Gaussians~\cite{li2024animatable}, and Relightable Gaussians~\cite{li2023animatable} learn pose-dependent Gaussian properties via texture-like maps over the body template. Vid2Avatar-Pro~\cite{guo2025vid2avatar} further introduces projective front/back maps and a pretrained universal prior for clothed avatar reconstruction. Other approaches employ CNN-based regressors in UV space to predict Gaussian attributes or geometry~\cite{jiang2025uv, kwon2024generalizable, Pang_2024_CVPR}. While these methods enable efficient and controllable animation, the learned representations remain tightly coupled to the template's parameterization and deformation space.

\begin{figure}
	\centering
	\includegraphics[width=\linewidth, , trim=0 0 0 0, clip]{
		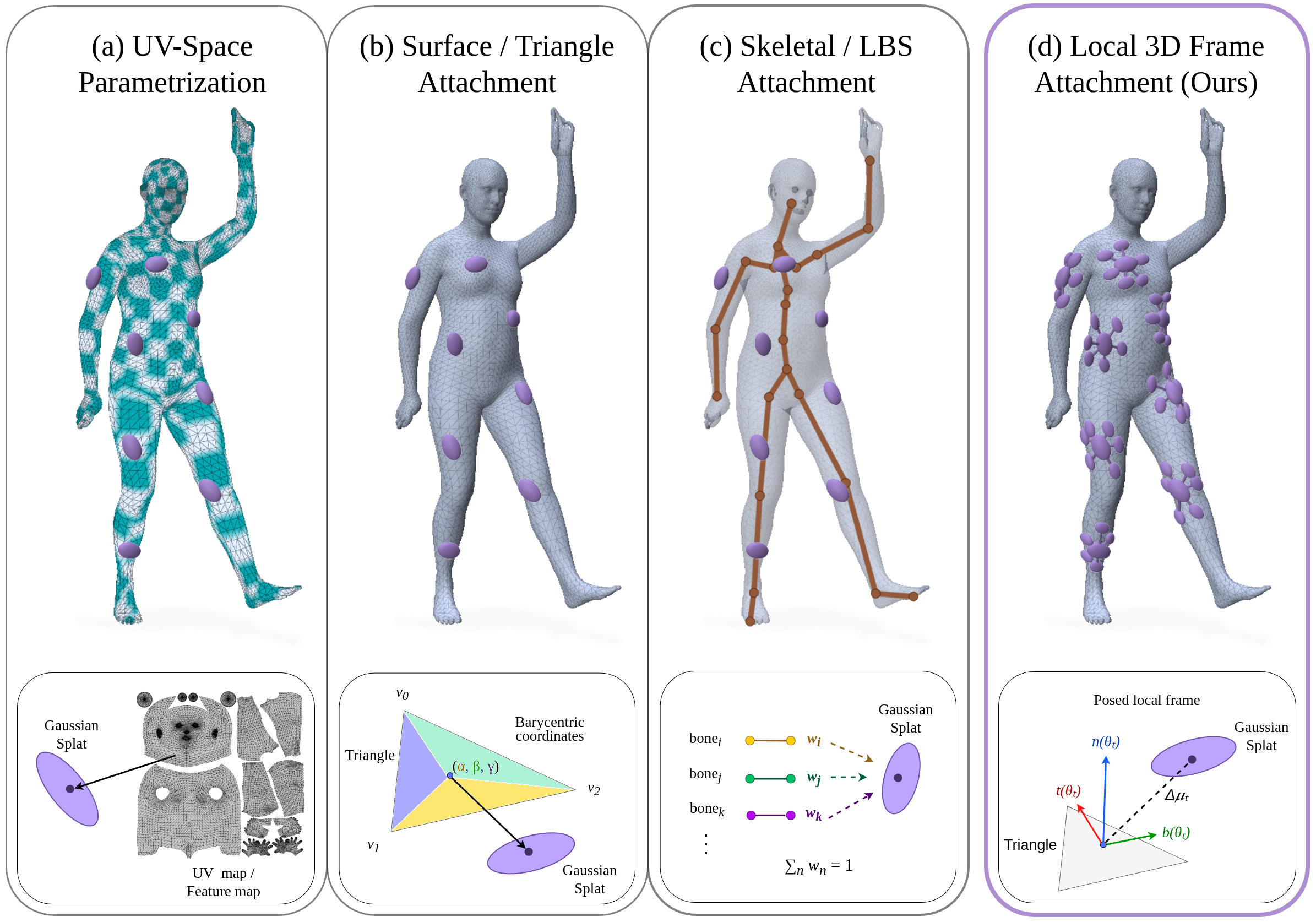
	} 
	\caption{Existing methods rely on UV maps, surface/triangle binding, or skeletal/LBS attachment, limiting deformation to the template. Our approach decouples anchors from template geometry, using canonical self-organization guided only by kinematic transport for flexible, non-rigid motion.}
	\label{fig:attachment}
\end{figure}

\begin{figure*}
    \centering
    \includegraphics[width=\linewidth, , trim=30 0 0 0, clip]{
        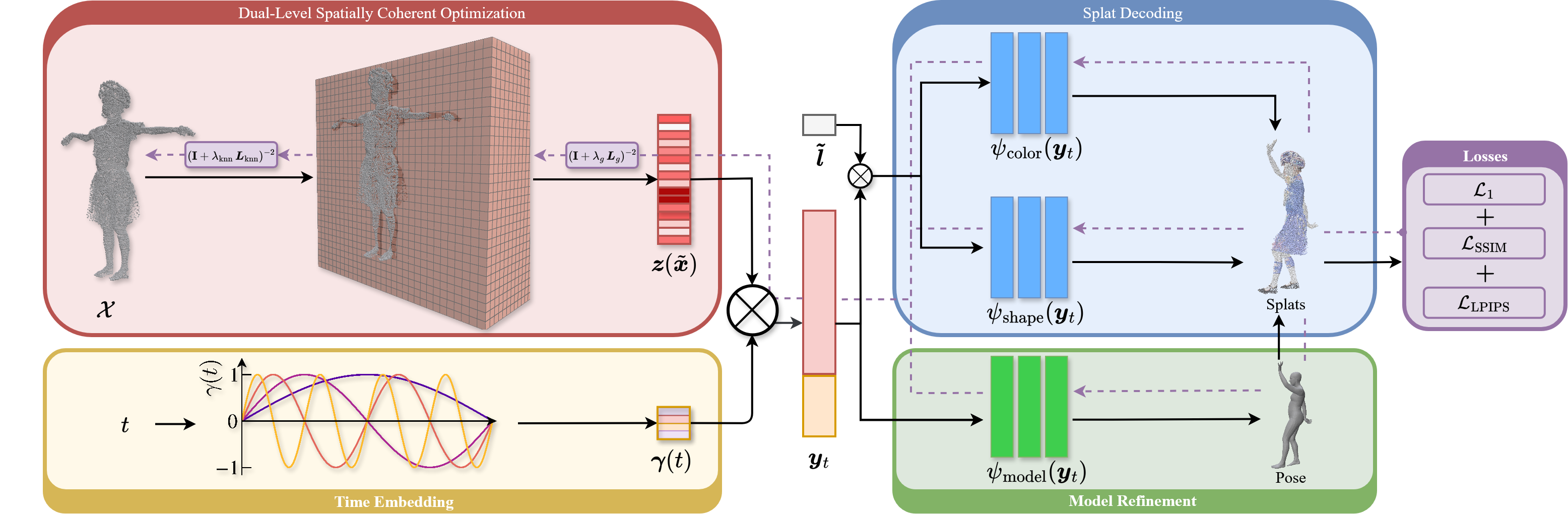
    } 
    \caption{Overview of PiG-Avatar. We learn a canonical, anchor-based Gaussian
    representation guided by a shared multi-resolution latent field. Conditioned
    on time and level of detail, lightweight decoders predict dynamic Gaussian attributes
    and refine a parametric proxy that transports anchors through time. This
    decoupled formulation enables stable geometry learning, articulated animation,
    and hierarchical reconstruction within a unified framework.
    }
    \label{fig:overview}
\end{figure*}

\subsection{Hybrid Mesh-Gaussian Representations}

Several recent methods combine Gaussian splatting with explicit mesh representations to leverage the advantages of both paradigms. These hybrid approaches typically embed Gaussian primitives directly on mesh surfaces and learn neural models to predict pose-dependent appearance or local deformations.

GoMAvatar~\cite{Wen_2024_CVPR}, RMAvatar~\cite{peng2025rmavatar}, and SplattingAvatar~\cite{shao2024splattingavatar} represent avatars using mesh-attached Gaussians combined with neural predictors for dynamic Gaussian attributes. Other works introduce layered or spatially distributed representations to better capture clothing deformation and local detail, including layered Gaussian avatars~\cite{lin2024layga} and interpolated spatial MLPs~\cite{zhan2025real}. Additional approaches integrate Gaussian rendering with implicit geometry~\cite{chen2024meshavatar,yuan2024gavatar}, physics-based garment simulation~\cite{lee2025mpmavatar}, or relightable rendering~\cite{choi2025relightable}.

Despite their strong rendering quality and efficiency, most existing Gaussian avatars remain fundamentally template-centered: Gaussian primitives are either attached to the template surface or directly parameterized in template space, thereby restricting the deformation to that of the model (see \Cref{fig:attachment}). In contrast, our approach learns a spatially adaptive canonical Gaussian
representation that is only kinematically guided by the articulated template,
allowing more flexible modeling of loose clothing and complex non-rigid motion
without requiring template-constrained geometry.

\section{Method}
\label{sec:method}
Given a multi-view video sequence consisting of images
$\{I_{c,t}\}$ captured from camera ${c \in \{1,\dots,C\}}$ at time
${t \in \{1,\dots,T\}}$, together with per-frame pose parameters obtained from
a parametric body model such as SMPL-X, our goal is to reconstruct a high-fidelity
animatable avatar that supports novel-view rendering and pose-driven animation.
Our formulation decouples articulated motion from canonical geometric representation, as illustrated in \Cref{fig:overview}.
We first use the body model purely as a kinematic prior to define temporally consistent
articulation (\Cref{sec:articulation}), and introduce a transport mechanism that associates
learnable Gaussian anchors with the articulated structure while allowing them to freely
organize in canonical 3D space (\Cref{sec:splat_attachment}). Subject geometry and appearance
are represented by these anchors together with a shared neural field parameterized as a
multi-resolution latent grid, which is optimized using a dual-level spatially coherent
strategy that regularizes both the neural field and the canonical anchor geometry
(\Cref{sec:dual-level}). Finally, time-dependent geometry and appearance properties of the
Gaussian primitives are decoded from the shared canonical representation using temporally
conditioned decoders (\Cref{sec:decoding}). Because all levels of detail share the same
underlying representation, the model naturally supports hierarchical reconstruction and
multi-resolution rendering within a unified framework.

\subsection{Anchor-based Gaussian Avatar Model}
\label{sec:articulation} 
\begin{figure}
    \centering
    \includegraphics[width=\linewidth, , trim=650 320 10 180, clip]{
        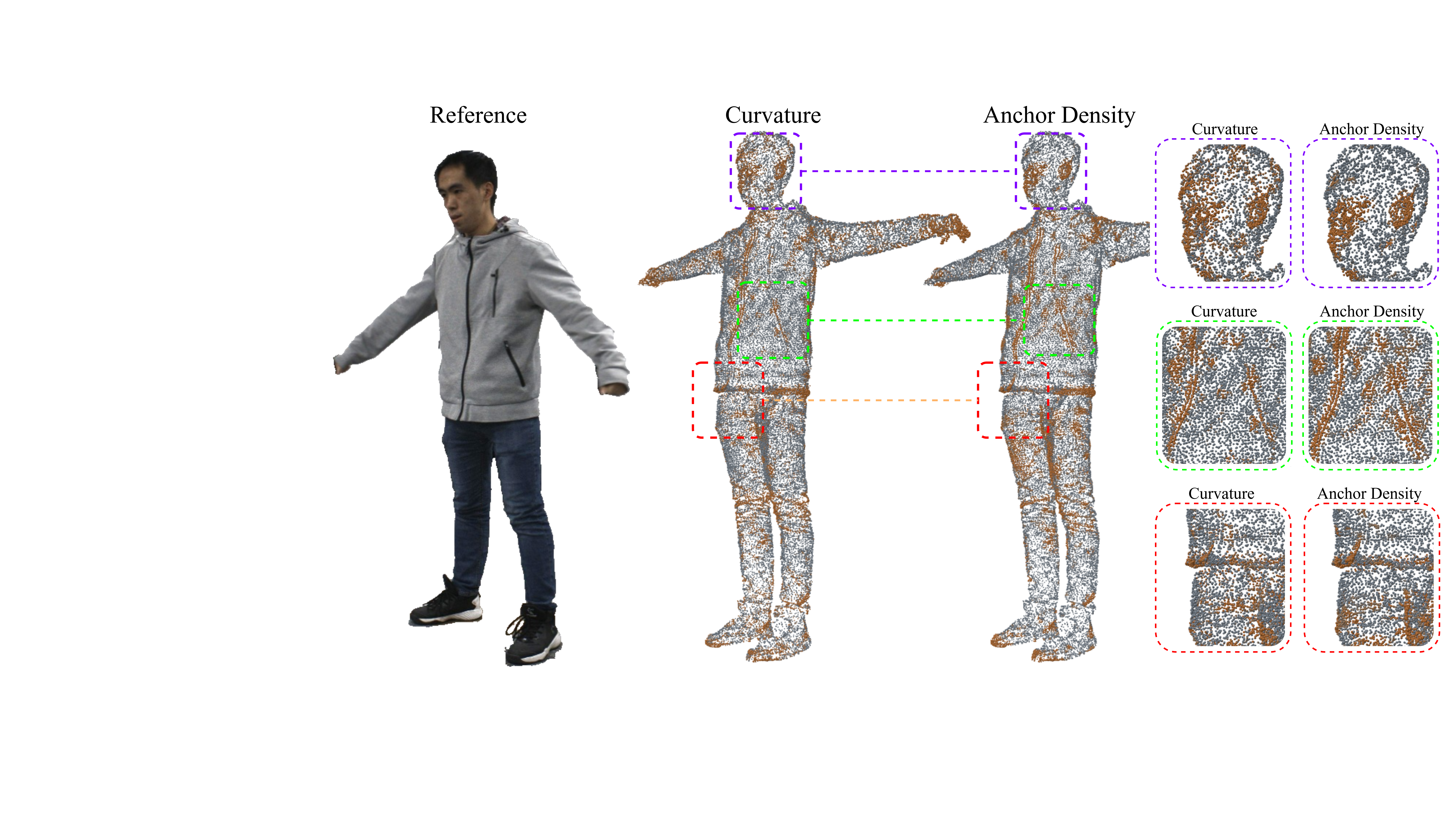
    } 
    \caption{Emergent anchor density from our spatially coherent optimization. Anchors concentrate in high-curvature regions, areas with strong appearance variation, and locations with complex local motion, while smoother, uniform regions remain sparse. 
    }
    \label{fig:anchor-density}
\end{figure}

We represent the avatar by a canonical set of 3D anchor points
${\set{X} = \{\vec{x}_{i}\}_{i=1}^{N}}$ that define the support of the Gaussian
representation. For each anchor $\vec{x}_{i}$, a spatial feature vector
${\vec{z}(\vec{x}_{i})}$ is queried from a continuous neural field in canonical
space, parameterized as a multi-resolution latent grid. This formulation decouples
feature storage from anchor placement, such that representational capacity is not
directly tied to the number or distribution of anchors. As a result, anchors can
adaptively reorganize toward regions requiring increased geometric or appearance
complexity, as depicted in \Cref{fig:anchor-density}. Temporal variation is introduced only at the decoding stage by conditioning
the queried canonical features on a time embedding. The canonical representation therefore
remains shared across the entire sequence, while temporally conditioned MLP decoders predict
dynamic Gaussian attributes
$\{\vec{\mu}_{i,t},\vec{q}_{i,t},\vec{s}_{i,t},o_{i,t}, \vec{c}_{i,t}\}$
as well as refined parametric-model parameters
$\{\vec{\beta}, \vec{\theta}_{t}\}$.

\paragraph{Nested Hierarchical Anchor Representation}
To support reconstruction at multiple levels of detail (LOD) within a unified
representation, we define a hierarchy of $ L $ nested anchor subsets. Starting from the
full set of canonical anchors at the finest level, we recursively construct coarser
levels by randomly subsampling anchors from the next finer level during initialization.
This produces a nested hierarchy
${\mathcal{X}_{1}\subset \dots \subset \mathcal{X}_{L} = \set{X}}$,
where each coarser level contains half as many anchors as the subsequent finer level,
i.e., ${\lvert \mathcal{X}_{l-1}\rvert = 0.5\, \lvert \mathcal{X}_{l}\rvert}$.
The sampled indices are fixed after initialization and consistently define the
active anchor subsets throughout training and inference.

This hierarchical construction couples optimization across levels of detail.
All anchor subsets query the same canonical neural field, while the proposed
dual-level spatially coherent optimization (see \Cref{sec:dual-level}) propagates
updates both through the shared field representation and across neighboring anchors.
Consequently, supervision applied at a coarse level also improves the canonical
features used by finer levels. Moreover, anchors that are inactive in the current
LOD continue to receive indirect geometric updates through neighborhood
coupling, allowing finer LODs to remain consistent with the optimized coarse
structure. In this way, all levels are jointly optimized within a single shared
hierarchical representation.

\paragraph{Canonical Neural Field}
We represent subject-specific canonical geometry and appearance using a shared
neural field defined over canonical space. To parameterize this field, we adopt
the preconditioned multi-resolution latent grid representation of
Neu-PiG \cite{kaltheuner2026neu}. Concretely, the canonical volume is covered by
$G$ learnable regular grids with progressively increasing spatial resolution.
Each grid stores a latent feature vector at every grid vertex. Given a normalized
canonical query position ${\tilde{\vec{x}}_i\in [-1,1]^{3}}$, we obtain a feature
${\vec{z}_{g}(\tilde{\vec{x}}_{i}) \in \mathbb{R}^d}$ from grid level $g$ via trilinear interpolation.
The final canonical feature descriptor is computed by averaging the interpolated
features across all grid levels:
\begin{equation}
    \vec{z}(\tilde{\vec{x}}_{i}) = \frac{1}{G}\sum_{g=1}^{G}\vec{z}_{g}(\tilde{\vec{x}}_{i}).
    \label{eq:latent_multires}
\end{equation}
This multi-resolution parameterization enables the field to represent structure
across different spatial scales while maintaining a fixed latent feature dimensionality.

\subsection{3D Barycentric Anchor Transport}
\label{sec:splat_attachment}

To enable temporally consistent articulation without restricting representational
flexibility, we associate each canonical anchor point $\vec{x}_{i}$ with the surface
of a parametric body model (here, SMPL-X), which serves solely as a kinematic transport
proxy throughout optimization. The body model does not define the final geometry of
the avatar; instead, it provides structured motion guidance, while the Gaussian
decoders (see \Cref{sec:decoding}) model local geometry and appearance around the transported
anchors. Although we use SMPL-X in our experiments, the formulation is compatible
with other articulated parametric geometry models.

\paragraph{Canonical Attachment}
Anchor binding is performed directly in 3D using local surface frames rather than a 2D UV parameterization. Unlike UV binding, which requires a bijective map from the avatar surface and cannot capture multi-layer geometry or large off-surface displacements, 3D local-frame attachment provides kinematic guidance without imposing such constraints. Each canonical anchor is initialized from a reference surface point $\vec{x}_{i}^{\mathrm{init}}$, sampled from the parametric model and stored as a fixed attachment. During optimization, the canonical anchor position $\vec{x}_{i}$ remains fully learnable and may deviate from its initial surface location, while the attachment parameters remain fixed, preserving temporally consistent transport under articulated deformation and allowing anchors to reorganize freely in canonical space.

\paragraph{Pose-Dependent Transport}
To transport an anchor $\vec{x}_{i}$ into a posed frame at time $t$, we express
its canonical offset relative to the reference surface point in the local canonical
surface frame, rotate this offset into the posed local frame induced by the deforming
proxy mesh, and add it to the posed surface position defined by the fixed barycentric
attachment. Given pose parameters $\vec{\theta}_{t}$, we evaluate the posed reference
surface point $\vec{x}_{i}^{\mathrm{init}}(\vec{\theta}_{t})$ on the deformed proxy
mesh, together with the corresponding
posed local frame
$(\vec{t}_{i}(\vec{\theta}_{t}), \vec{b}_{i}(\vec{\theta}_{t}),
\vec{n}_{i}(\vec{\theta}_{t}))$.
The relative rotation between the canonical and posed local frames is defined as
\begin{equation}
    \vec{R}_{i,t}= \Big[ \vec{t}_{i}(\vec{\theta}_{t}) \,\Big\vert\, \vec{b}_{i}(
    \vec{\theta}_{t}) \,\Big\vert\, \vec{n}_{i}(\vec{\theta}_{t}) \Big]
    \Big [ \vec{t}_{i}\,\Big\vert\, \vec{b}_{i}\,\Big\vert\, \vec{n}_{i}\Big]^{\top}.
    \label{eq:transport_rotation}
\end{equation}
Using this rotation, the posed anchor position is computed as
\begin{equation}
    \vec{x}_{i}(\vec{\theta}_{t}) =
    \vec{x}_{i}^{\mathrm{init}}(\vec{\theta}_{t})
    + \vec{R}_{i,t}\,
    \Big(\vec{x}_{i}- \vec{x}_{i}^{\mathrm{init}}\Big).
    \label{eq:transported_anchor}
\end{equation}
This formulation keeps the barycentric attachment fixed while consistently rotating
the canonical anchor offset with the evolving local surface frame under articulated
deformation. As a result, anchors inherit temporally coherent articulated motion
while remaining unconstrained by the topology and geometry of the proxy surface.

\subsection{Dual-Level Spatially Coherent Optimization}
\label{sec:dual-level}
Maintaining kinematic coherence for freely moving Gaussian anchors is essential to producing stable temporal correspondences. However, naive regularization can inadvertently restrict the model’s expressive capacity. To address this, we propose a dual-level spatially coherent preconditioning scheme, which combines Sobolev-preconditioned updates for the canonical neural field with a KNN-based preconditioning of the anchor positions. This design ensures both the latent field and the explicit anchor geometry evolve coherently in space while preserving the flexibility required to model complex, layered clothing and non-rigid deformations.

\paragraph{Grid-based Field Preconditioning}
For each latent grid level $g$, the gradient with respect to the grid parameters $\vec{z}_{g}$ is filtered using a resolution-dependent Laplacian operator before updating the grid:
\begin{equation}
    \vec{z}_{g}\leftarrow \vec{z}_{g}- \eta_{g}\, \left(\vec{I}+\lambda_{g}\,\mat
    {L}_{g}\right )^{-2}\, \frac{\partial \mathcal{L}}{\partial \vec{z}_{g}}, \label{eq:sobolev_update}
\end{equation}
where $\eta_{g}$ is the learning rate, $\lambda_g$ controls the smoothing strength at level $g$, and $\mat{L}_g$ encodes local neighborhood connectivity~\cite{kaltheuner2026neu}. This Sobolev preconditioning acts as a low-pass filter on the latent updates, promoting smooth propagation of supervision across neighboring regions. As a result, nearby anchors querying the field obtain consistent descriptors, stabilizing the learning dynamics and improving temporal coherence.

\paragraph{KNN-based Anchor Preconditioning}
In parallel, we regularize anchor positions using neighborhood-aware preconditioning defined on a fixed KNN graph in canonical space. Denoting the graph Laplacian as $\mat{L}_{\mathrm{knn}}$, anchor positions are updated as:
\begin{equation}
    \vec{x}\leftarrow \vec{x}- \eta_{\mathrm{knn}}\, \left(\vec{I}+\lambda_{\mathrm{knn}}
    \,\vec{L}_{\mathrm{knn}}\right )^{-2}\, \frac{\partial \mathcal{L}}{\partial
    \vec{x}}. \label{eq:anchor_precond_update}
\end{equation}
Here, $\eta_{\mathrm{knn}}$ is the anchor learning rate, and $\lambda_{\mathrm{knn}}$ governs the spatial coupling strength. These neighborhood relations are defined once in the canonical space and remain fixed throughout optimization, ensuring that nearby anchors remain coupled even after transport through the posed model. This coupling encourages local structural coherence, prevents anchor drift, and is particularly effective for modeling loose garments, folds, and other non-rigid motions.

Together, the Sobolev and KNN preconditionings induce an emergent self-organization of the anchors, allowing them to migrate naturally toward regions of high curvature, complex motion, or appearance variation without relying on hand-crafted heuristics. This emergent behavior improves spatial coverage of critical regions while preserving temporal consistency. Furthermore, this canonical coupling naturally supports hierarchical level-of-detail reconstruction: even inactive anchors in a given forward pass continue to receive indirect updates through their neighbors, reinforcing coherent structure across resolutions.

\begin{figure}
    \centering
    \includegraphics[width=\linewidth, , trim=0 0 0 0, clip]{
        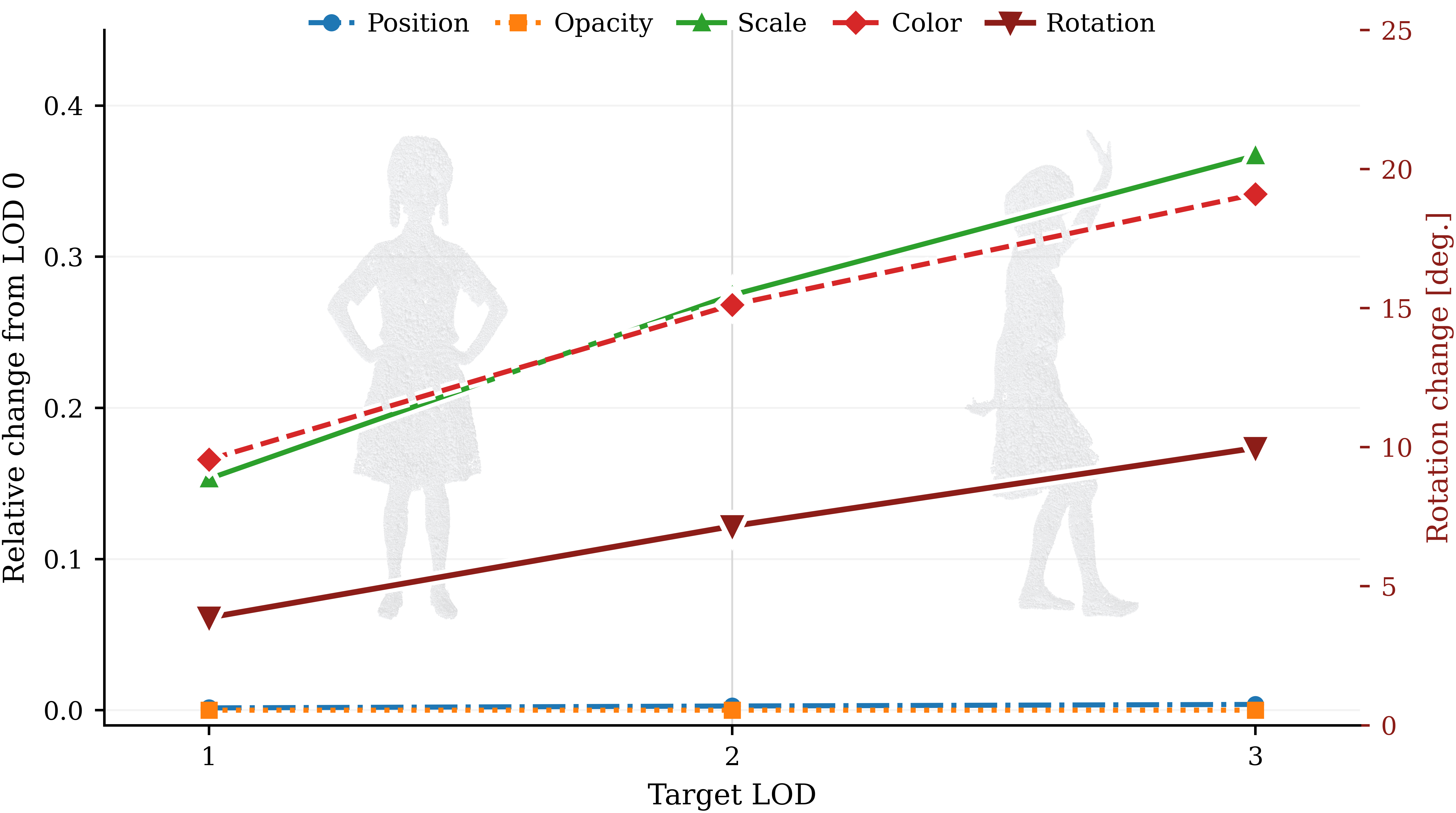
    } 
    \caption{ Position and opacity remain stable across target LODs, whereas scale and color
change substantially with increasing detail. Rotation varies more moderately,
reaching about $10^\circ$ at the highest target LOD. Overall, LOD conditioning
primarily affects splat extent and appearance while preserving stable anchor
positions across nested subsets.}
    \label{fig:lod}
\end{figure}

\subsection{Time-Conditioned Gaussian Parameter Decoding}
\label{sec:decoding}

To model temporal variation, we combine the anchor-wise canonical descriptors with a temporal embedding and level-of-detail encoding.

For time, we adopt a Fourier feature embedding~\cite{tancik2020fourier} of the normalized timestep, ${\tilde{t}=(t-1)/(T-1) \in [0,1]}$, mapping it to a multi-frequency sinusoidal representation:
\begin{equation}
    \vec{\gamma}(t) = \left[\sin(\pi \nu_f \tilde{t}), \cos(\pi \nu_f \tilde{t})\right]_{f=1}^{F}, \label{eq:time_embedding}
\end{equation}
where the exponentially increasing frequencies are ${\nu_f = 2^{f-1}}$.
Following Neu-PiG~\cite{kaltheuner2026neu}, this embedding is used to modulate anchor-wise descriptors for time-dependent decoding, enabling fast and slow temporal variations to be represented within the same latent space.

Choosing $F$ too small limits the temporal bandwidth and prevents accurate modeling of rapid pose- or appearance-dependent changes, while excessively large $F$ allocates capacity to unnecessarily fine temporal variations. We therefore scale the number of frequency bands with the sequence length, setting ${F = \lceil \log_2 T \rceil}$. This produces a $2F$-dimensional temporal descriptor suitable for both slow and fast temporal dynamics without fixing the embedding size a priori.

For hierarchical reconstruction, we further append the normalized level-of-detail variable ${\tilde{l}=(l-1)/(L-1) \in [0,1]}$ to the decoder input. The final conditioning vector for an anchor $i$ at timestep $t$ and LOD $l$ is:
\begin{equation}
    \vec{y}_{i,t,l} = \left[ \vec{z}(\tilde{\vec{x}}_i), \vec{\gamma}(t), \tilde{l} \right] \in \mathbb{R}^{d + 2F + 1}. \label{eq:decoder_input}
\end{equation}
As also shown in \Cref{fig:lod}, this formulation ensures that temporal dynamics and multi-resolution effects are naturally incorporated into the anchor-wise decoding of both geometry and appearance.

\subsubsection{Appearance Decoder}
The view-dependent appearance of each anchor is predicted via a decoder $\Psi_{\mathrm{color}}$ mapping $\vec{y}_{i,t,l}$ to spherical harmonics coefficients:
\begin{equation}
    \vec{c}_{i,t}= \Psi_{\mathrm{color}}(\vec{y}_{i,t,l}). \label{eq:color_decoder}
\end{equation}
For readability, we omit the explicit dependence of the predicted attributes on $l$ in the notation; it is implicitly given through the decoder input $\vec{y}_{i,t,l}$.

We use separate MLP decoders for geometry and appearance to allow each to specialize to distinct spatial-frequency characteristics. In articulated human avatars, smooth surfaces like skin or clothing may still contain high-frequency details from textures, wrinkles, or fabric patterns, while regions with complex geometry, such as folds or facial features, can remain relatively uniform in color. By disentangling the decoding, each decoder focuses on the structural properties of its domain, improving representational capacity and enabling more accurate, high-fidelity reconstruction.

\subsubsection{Shape Decoder}
Geometric attributes of each Gaussian anchor are predicted by a dedicated decoder $\Psi_{\mathrm{shape}}$:
\begin{equation}
    \{\Delta \vec{\mu}_{i,t},\Delta \vec{q}_{i,t},\vec{s}_{i,t},o_{i,t}\} = \Psi_{\mathrm{shape}}
    (\vec{y}_{i,t,l}), \label{eq:shape_decoder}
\end{equation}
where ${\Delta \vec{\mu}_{i,t} \in \mathbb{R}^3}$ and ${\Delta \vec{q}_{i,t} \in \mathbb{R}^4}$ denote residual position and rotation offsets, ${\vec{s}_{i,t} \in \mathbb{R}^3}$ represents Gaussian scales, and ${o_{i,t} \in \mathbb{R}}$ encodes opacity. To ensure valid outputs, we adopt task-specific parameterizations: rotations are expressed as unit quaternions, scales are predicted in log-space, and opacities are constrained to $(0,1)$.

The residual terms $\Delta \vec{\mu}_{i,t}$ and $\Delta \vec{q}_{i,t}$ capture local dynamic effects not explained by skeletal transport alone, such as clothing motion, folds, and subtle deformations. These residuals are combined with the pose-dependent transport described in \Cref{sec:splat_attachment} to compute final splat positions and orientations. Specifically, the final splat center for a posed anchor $\vec{x}_{i}(\vec{\theta}_t)$ from the parametric model is
\begin{equation}
    \vec{\mu}_{i,t}= \vec{x}_{i}(\vec{\theta}_{t}) + \Delta \vec{\mu}_{i,t}, \label{eq:splat_position}
\end{equation}
and the residual rotation, applied in the local posed frame, yields
\begin{equation}
    \vec{q}_{i,t}= \vec{q}(\vec{R}_{i,t})\,\Delta \vec{q}_{i,t}, \label{eq:splat_rotation}
\end{equation}
where $\vec{q}(\vec{R}_{i,t})$ converts the transported rotation $\vec{R}_{i,t}$ to quaternion space. This formulation preserves kinematic coherence while allowing anchors to deviate freely from the template, enabling the representation of complex layered and off-body clothing.

\begin{figure}
    \centering
    \includegraphics[width=\linewidth, , trim=580 300 700 200, clip]{
        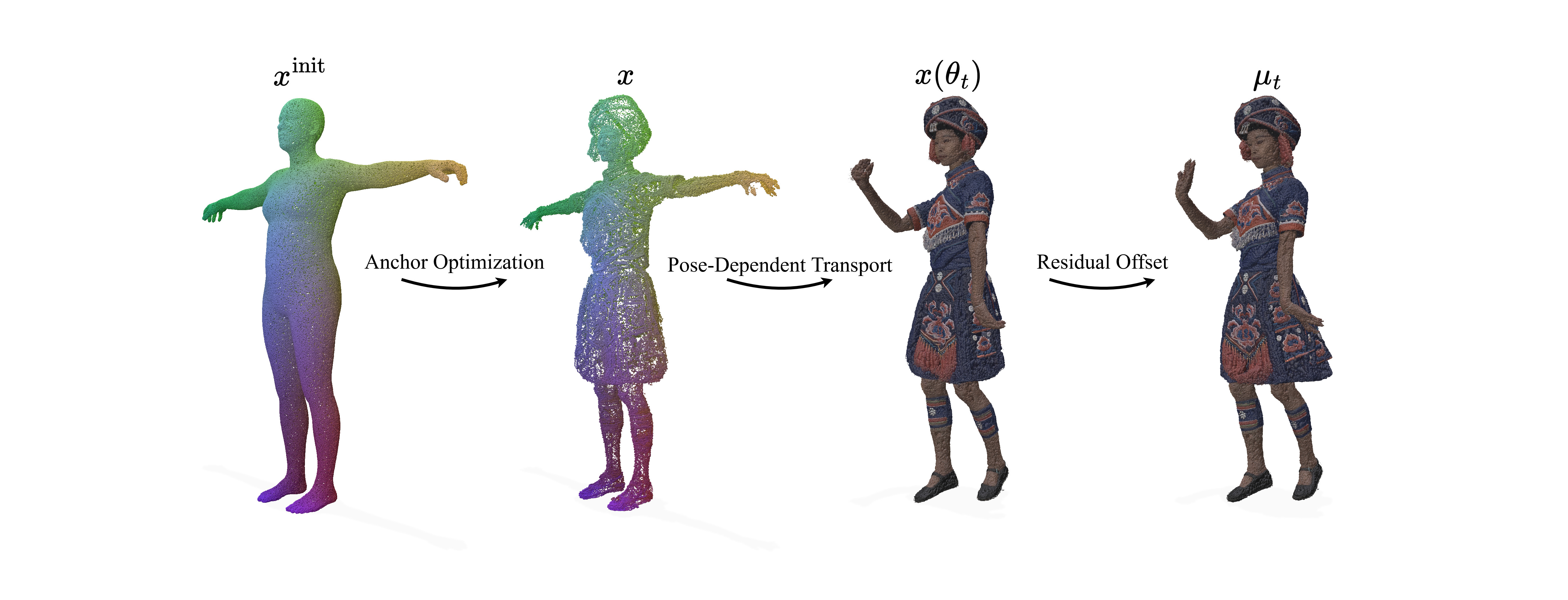
    } 
    \caption{ Illustration of anchor transport through the deforming proxy mesh. Anchors first self-organize in canonical space, then follow kinematic motion with local offsets, producing the final Gaussian centers that track surface deformation consistently. }
    \label{fig:transport}
\end{figure}

\subsubsection{Global Parametric-Model Refinement}
In addition to anchor-wise Gaussian attributes, we refine the underlying parametric body model to improve motion alignment and kinematic coherence. We predict residual updates to the model parameters using a global descriptor $\bar{\vec{z}}$ extracted from the canonical neural field. While Gaussian decoders operate on local anchor features, the parametric model encodes sequence-level structure and motion, motivating the use of a single global latent feature, which is obtained by averaging canonical latent features across all anchors:
\begin{equation}
    \vec{y}_{t} = [\bar{\vec{z}}, \vec{\gamma}(t)],
    \qquad
    \bar{\vec{z}} = \frac{1}{N}\sum_{i=1}^{N}\vec{z}(\tilde{\vec{x}}_{i}). \label{eq:model_input}
\end{equation}
Unlike local decoders, this branch does not depend on the reconstruction level $l$, as it captures sequence-wide motion and shape information.

Given an initial parametric body model $(\vec{\beta}^{\mathrm{init}}, \vec{\theta}^{\mathrm{init}}_t)$, we predict residual updates using a single MLP, $\Psi_{\mathrm{model}}$:
\begin{equation}
    \{\Delta \vec{\beta}_{t}, \Delta \vec{\theta}_{t}\} = \Psi_{\mathrm{model}}(\vec
    {y}_{t}), \label{eq:model_decoder}
\end{equation}
where $\Delta \vec{\beta}_t$ and $\Delta \vec{\theta}_t$ denote shape and pose corrections, respectively. Since the shape parameters $\vec{\beta}$ are time-invariant, we aggregate the predicted shape residuals across all timesteps during training to produce a single sequence-level update, $\Delta \vec{\beta}$. This averaging couples gradients temporally, producing a stable estimate of the overall shape. Pose corrections $\Delta \vec{\theta}_t$ remain timestep-specific, adapting to local motion, but are refined to preserve the articulated structure across the sequence. These globally-consistent corrections are important for properly disentangling anchor residuals and enabling the 3D barycentric anchor transport to establish stable, temporally consistent correspondences across frames.

The final model parameters are therefore
\begin{equation}
    \vec{\beta}= \vec{\beta}^{\mathrm{init}}+ \Delta \vec{\beta}, \qquad \vec{\theta}
    _{t}= \vec{\theta}^{\mathrm{init}}_{t}+ \Delta \vec{\theta}_{t}. \label{eq:model_final}
\end{equation}
By predicting residuals relative to the initialization, rather than absolute values, this formulation stabilizes optimization and maintains the kinematic coherence used to guide anchor transport.

\subsection{Optimization Objectives}
\label{sec:optimization}

We supervise reconstruction entirely in the image domain. Following MMLPHuman~\cite{zhan2025real}, all losses are applied with silhouette-aware masking to suppress ambiguous boundary pixels. The overall objective is
\begin{equation}
    \label{eq:loss}\mathcal{L}= \lambda_{1}\, \mathcal{L}_{1}+ \lambda_{\text{SSIM}}
    \,\mathcal{L}_{\text{SSIM}}+ \lambda_{\text{LPIPS}}\, \mathcal{L}_{\text{LPIPS}},
\end{equation}
where $\mathcal{L}_1$, $\mathcal{L}_{\text{SSIM}}$, and $\mathcal{L}_{\text{LPIPS}}$ denote pixel-wise $L_1$, structural similarity, and perceptual losses, weighted by ${\lambda_1=8}$, ${\lambda_{\text{SSIM}}=2}$, and ${\lambda_{\text{LPIPS}}=2}$, respectively.

In contrast to many dynamic avatar methods, we do not employ explicit geometric or temporal regularization. Spatial coherence emerges naturally through the dual-level preconditioning scheme, while kinematic consistency and stable correspondences are maintained by the structured parametric model and 3D barycentric anchor transport, without additional temporal penalties.

For hierarchical reconstruction, a level-of-detail $l$ is randomly sampled at each training iteration, and losses are computed only on the corresponding anchor subset $\mathcal{X}_l$, allowing coarse-level supervision to propagate to finer resolutions.

\subsection{Implementation Details}

We represent the avatar with a multi-resolution latent grid of ${G=6}$ levels and ${d=16}$ features per cell. The coarsest level has a resolution of $8^3$, and each finer level increases the resolution by a factor of 16 along each axis, reaching a finest resolution of $88^3$. The Gaussian splat representation is initialized with $100{,}000$ anchors, without additional densification or pruning. View-dependent color is modeled using spherical harmonics of degree 3.

The model is optimized end-to-end with Adam~\cite{kingma2015adam} using the \texttt{gsplat} renderer. The parametric model network $\Psi_{\mathrm{model}}$ has three hidden layers of 32 units each, while the color and shape decoders, $\Psi_{\mathrm{color}}$ and $\Psi_{\mathrm{shape}}$, each use five hidden layers of 512 units. Learning rates are $10^{-4}$ for $\Psi_{\mathrm{model}}$ and $10^{-3}$ for both decoders. For the latent grid, the base learning rate is $10^{-2}$ at the coarsest level and increases by $1.5\times$ per finer level; grid smoothness is scaled similarly, starting at ${\lambda_g=2}$ and increasing by $1.25\times$ per level. Anchor positions use a learning rate of $2\cdot10^{-5}$ with ${\lambda_{\text{knn}}=8}$.

Training is performed for 100{,}000 iterations, sampling 4 random views and 5 timesteps per iteration.

\section{Evaluation}

\subsection{Datasets}
We evaluate PiG-Avatar on THuman4~\cite{zheng2022structuredlocalradiancefields} and DNA-Rendering~\cite{cheng2023dnarenderingdiverseneuralactor}, following established protocols to ensure fair comparison with prior work.
\begin{itemize}[leftmargin=*]
    \item \emph{THuman4}
contains dynamic performances of clothed human subjects captured with 24 calibrated RGB cameras at 30\,FPS and a resolution of ${1330\times1150}$. We evaluate on all three available subjects. For novel-view evaluation, 23 cameras are used for training while one camera is held out. For novel-pose evaluation, we train on the first 2000 timesteps and test on the remaining 500 timesteps.

    \item \emph{DNA-Rendering}
features multi-view sequences captured with 60 synchronized cameras and exhibits complex garments, loose clothing, and challenging non-rigid motion. We evaluate on the sequences \texttt{0165}, \texttt{0166}, and \texttt{0206}. For novel-view evaluation, 48 cameras are used for training and 12 cameras are held out. For novel-pose evaluation, we train on 80\% of each sequence and test on the remaining 20\% of frames.
\end{itemize}
For novel-pose evaluation across both datasets, each query pose is mapped to the closest captured timestep in SMPL-X body-pose space. We compute the sum of per-joint geodesic distances in $\mathrm{SO}(3)$ between joint rotations and select the nearest timestep as the reference clothing state.

\begin{table}[t]
    \centering
    \caption{Quantitative comparison on THuman4 across novel views, and novel poses.
    PSNR/SSIM $\uparrow$, LPIPS/FID $\downarrow$. Red and orange boxes indicate best and second-best results, respectively.}
    \label{tab:thuman_results}
    \renewcommand{\arraystretch}{1.1}
    \setlength{\tabcolsep}{3pt}
    \resizebox{\linewidth}{!}{%
    \begin{tabular}{l|cccc|cccc|c}
        \toprule                              & \multicolumn{4}{c}{Novel Views} & \multicolumn{4}{c|}{Novel Poses} & Splats\\
        \cmidrule(lr){2-5} \cmidrule(lr){6-9} & PSNR$\uparrow$                  & SSIM$\uparrow$                 & LPIPS$\downarrow$ & FID$\downarrow$ & PSNR$\uparrow$ & SSIM$\uparrow$ & LPIPS$\downarrow$ & FID$\downarrow$ &  \\
        \midrule
        PoseVocab                             & 24.90                           & 0.965                          & 0.038             & 109.72          & 24.06          & 0.958          & 0.043             & 105.29          & - \\
        AnimatableGS               & 28.25                           & 0.973                          & 0.036             & \phantom{0}56.83           & 27.38          & 0.961          & 0.043             & \phantom{0}42.49           & 361k \\
        MMLPHuman                             & \colorbox{best}{32.73}                           & \colorbox{best}{0.984}                          & \colorbox{best}{0.020}              & \phantom{00}\colorbox{second}{9.11}            & \colorbox{best}{29.37}          & \colorbox{best}{0.971}          & \colorbox{best}{0.029}             & \phantom{00}\colorbox{best}{7.14}            & \colorbox{second}{200k} \\
        \midrule Ours                & \colorbox{second}{32.57}                  & \colorbox{second}{0.982}                 & \colorbox{second}{0.023}    & \phantom{00}\colorbox{best}{8.07}   & \colorbox{second}{28.08} & \colorbox{second}{0.967} & \colorbox{second}{0.033}    & \phantom{0}\colorbox{second}{10.95}  & \colorbox{best}{100k} \\
        \bottomrule
    \end{tabular}%
    }
\end{table}

\begin{table}[t]
    \centering
    \caption{Quantitative comparison on DNA across training views, novel views,
    and novel poses. PSNR/SSIM $\uparrow$, LPIPS/FID $\downarrow$. Red and orange boxes indicate best and second-best results, respectively.}
    \label{tab:dna_results}
    \renewcommand{\arraystretch}{1.1}
    \setlength{\tabcolsep}{3pt}
    \resizebox{\linewidth}{!}{%
    \begin{tabular}{l|cccc|cccc|c}
        \toprule                              & \multicolumn{4}{c}{Novel Views} & \multicolumn{4}{c}{Novel Poses} & Splats\\
        \cmidrule(lr){2-5} \cmidrule(lr){6-9} & PSNR$\uparrow$                  & SSIM$\uparrow$                 & LPIPS$\downarrow$ & FID$\downarrow$ & PSNR$\uparrow$ & SSIM$\uparrow$ & LPIPS$\downarrow$ & FID$\downarrow$ & \\
        \midrule
        PoseVocab                             &   25.35                              &0.964                                &0.046                   &126.89                 &            23.64    &           \colorbox{best}{0.951}     &          0.057         &     146.48            & - \\
        AnimatableGS               &     28.33                            &0.960                                &0.037                   &216.72                 &24.22                &0.936                &0.052                   &223.73                 & 459k \\
        MMLPHuman                             &\colorbox{best}{31.83}                                 &\colorbox{second}{0.981}                                &\colorbox{best}{0.027}                   &\phantom{0}\colorbox{best}{13.18}                 &\colorbox{best}{25.58}                &   \colorbox{best}{0.951}             &\colorbox{best}{0.048}                   &\phantom{0}\colorbox{second}{36.13}                 & \colorbox{second}{200k} \\
        \midrule Ours                & \colorbox{second}{31.29}                  & \colorbox{best}{0.982}                 & \colorbox{second}{0.031}    & \phantom{0}\colorbox{second}{13.20}   & \colorbox{second}{25.12}      & \colorbox{second}{0.948}      & \colorbox{second}{0.051}         & \phantom{0}\colorbox{best}{32.45}       & \colorbox{best}{100k} \\
        \bottomrule
    \end{tabular}%
    }
\end{table}


\begin{figure}
    \centering
    \includegraphics[width=\linewidth, , trim=0 0 0 0, clip]{
        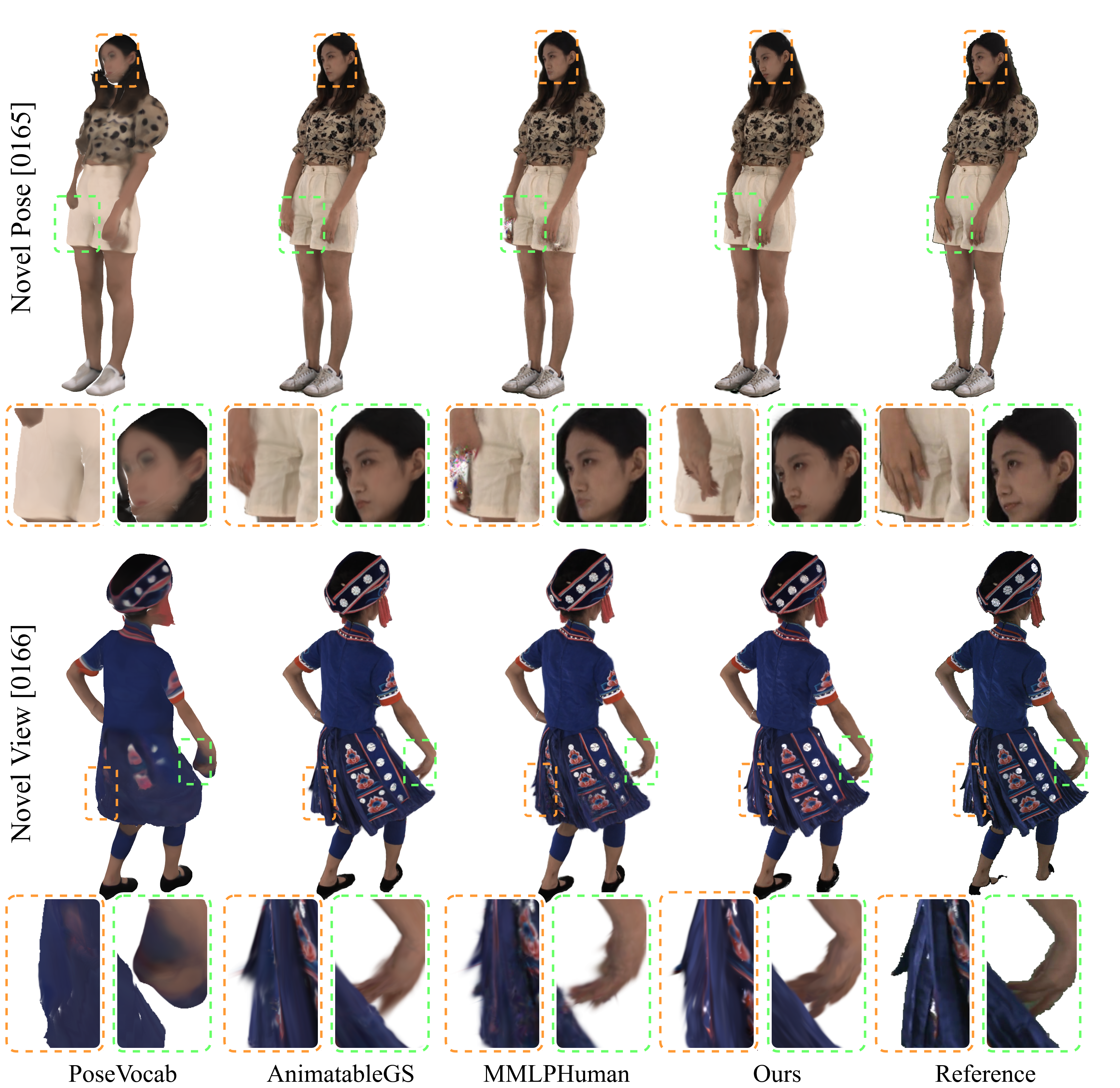
    }
    \caption{Qualitative comparison on DNA for novel-pose synthesis (top, 0165) and novel-view synthesis (bottom, 0166). Zoom-ins highlight differences in facial detail, hand geometry, and clothing reconstruction.}
    \label{fig:eval}
\end{figure}

\subsection{Baselines and Evaluation Metrics}
We compare PiG-Avatar against three state-of-the-art animatable human avatar methods: PoseVocab~\cite{li2023posevocab}, AnimatableGS~\cite{li2024animatable}, and MMLPHuman~\cite{zhan2025real}.

To evaluate reconstruction fidelity and perceptual quality, we adopt PSNR, SSIM~\cite{wang2004image}, LPIPS~\cite{zhang2018unreasonableeffectivenessdeepfeatures}, and FID~\cite{heusel2018ganstrainedtimescaleupdate} as standard image-based metrics commonly used in human avatar rendering:
PSNR and SSIM measure pixel-level reconstruction fidelity and structural similarity.
LPIPS reflects perceptual similarity as judged by human observers.
FID compares distributions of rendered and ground-truth images to assess perceptual realism.

\subsection{Quantitative and Qualitative Evaluation}
\label{sec:quantitative_qualitative}

\Cref{tab:thuman_results,tab:dna_results} compare PiG-Avatar to recent avatar reconstruction methods on THuman4 and DNA-Rendering. Our method achieves competitive reconstruction quality while using a topology-decoupled canonical representation and, unlike the baselines, learns a single hierarchical model that supports multiple levels of detail. On THuman4, PiG-Avatar matches MMLPHuman in novel-view quality, attains the best FID, and uses only half as many splats. For novel-pose evaluation, it outperforms AnimateableGS and PoseVocab and remains close to MMLPHuman despite the reduced number of primitives. On DNA-Rendering, PiG-Avatar maintains competitive novel-view performance and achieves the best FID for novel poses, demonstrating strong perceptual consistency under challenging pose generalization. Qualitative results in \Cref{fig:eval} show that PiG-Avatar preserves overall body shape, garment structure, and fine local appearance details in challenging regions such as faces, silhouettes, and textured clothing.

\subsection{Level-of-Detail Evaluation}
\label{sec:lod_evaluation}

\begin{figure}
    \centering
    \includegraphics[width=\linewidth, , trim=0 0 0 0, clip]{
        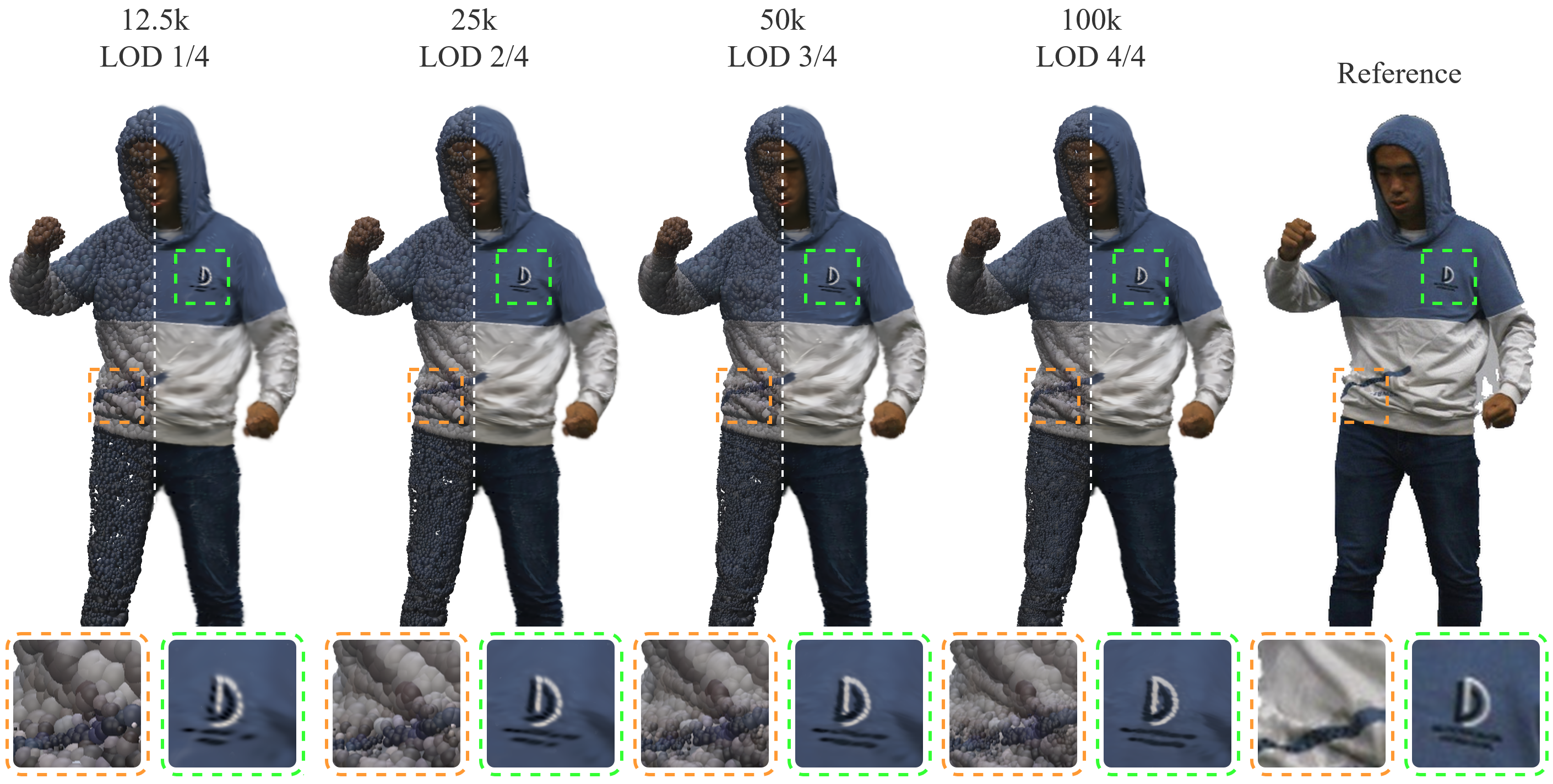
    }
    \caption{ LOD comparison of our shared hierarchical representation, showing
    splat structure colored by base color and corresponding renderings. Zoom-ins
    highlight local geometry and appearance details across resolutions. }
    \label{fig:lod_lvls}
\end{figure}

\Cref{tab:run_metrics} evaluates the level-of-detail behavior of our shared hierarchical representation. Reconstruction quality remains stable across resolutions while memory usage and runtime scale proportionally with the number of active splats. Even the coarsest level with $12.5$k splats achieves high-quality reconstruction and renders at $242$ FPS, whereas the full $100$k-splat representation further improves perceptual fidelity while maintaining real-time performance. \Cref{fig:lod_lvls} illustrates the corresponding visual trade-off, with coarse LODs preserving the global avatar structure and appearance and finer levels recovering additional local details in both the splat distribution and renderings. These results demonstrate progressive real-time rendering from a single trained model without requiring separate fixed-resolution avatars.

\begin{table}[t]
    \centering
    \caption{Memory consumption and rendering performance (FPS) across levels of
    detail (LOD), evaluated on novel views on the THuman4D dataset on an NVIDIA RTX
    5090 GPU. Red and orange boxes indicate best and second-best results, respectively.}
    \label{tab:run_metrics}
    \renewcommand{\arraystretch}{1.1}
    \resizebox{0.85\linewidth}{!}{%
    \begin{tabular}{cc|cccc|c|c}
    \toprule
        LOD          & Splats & PSNR$\uparrow$ & SSIM$\uparrow$ & LPIPS$\downarrow$ & FID$\downarrow$ & MB$\downarrow$ & FPS $\uparrow$ \\
        \midrule1/4  & 12.5k\,  & 32.52          & \colorbox{second}{0.981}          & 0.029             & 9.87            & \phantom{0}\colorbox{best}{8.68}          & \colorbox{best}{242}            \\
        2/4          & \phantom{0}25k  & 32.63          & \colorbox{second}{0.981}          & 0.027             & 8.99            & \phantom{0}\colorbox{second}{9.73}          & \colorbox{second}{159}            \\
        3/4          & \phantom{0}50k  & \colorbox{second}{32.64}          & \colorbox{best}{0.982} & \colorbox{second}{0.025}             & \colorbox{second}{8.34}            & 11.82          & \phantom{0}93             \\
        4/4          & 100k & 32.57          & \colorbox{best}{0.982} & \colorbox{best}{0.023}    & \colorbox{best}{8.07}   & 16.02          & \phantom{0}52             \\
        \midrule 1/1 & 100k & \colorbox{best}{32.66} & \colorbox{best}{0.982} & \colorbox{best}{0.023}    & 8.42            & 16.02          & \phantom{0}52             \\
        \bottomrule
    \end{tabular}%
    }
\end{table}


\subsection{Ablation Studies}
Our ablation study investigates the contribution of the main structural components of PiG-Avatar and analyzes its robustness to noise in the input SMPL-X parameters.

\paragraph{Structural Ablations}
To evaluate the contribution of PiG-Avatar's key structural components, we perform an ablation study in which individual elements are removed while all other settings remain unchanged, with results summarized in Table~\ref{tab:ablation_structural}. We test the effect of fixing canonical anchors without refinement to assess the role of anchor mobility, removing per-timestep splat offsets $\Delta \vec{\mu}_{i,t}$ to measure the impact of local dynamic adjustments, disabling dual-level preconditioning to examine its influence on spatially coherent canonical geometry and optimization stability, and replacing separate MLPs for splat shape and color with a single shared MLP to evaluate the benefit of decoupling geometry and appearance.

Removing the residual position offsets causes the largest performance drop, confirming that decoder-predicted local dynamics are the most critical component for modeling non-rigid clothing motion. Notably, even without anchor optimization, the model still achieves the second-best overall results, indicating that the preconditioned shared latent representation already provides a strong geometric prior when decoded by the MLPs. In contrast, removing dual-level preconditioning leads to clear degradation across all metrics, highlighting its importance for spatially coherent canonical geometry and stable optimization. Using a single splat MLP also reduces performance, showing that disentangling shape and appearance remains beneficial.

\begin{figure}
    \centering
    \includegraphics[width=\linewidth, , trim=300 100 80 150, clip]{
        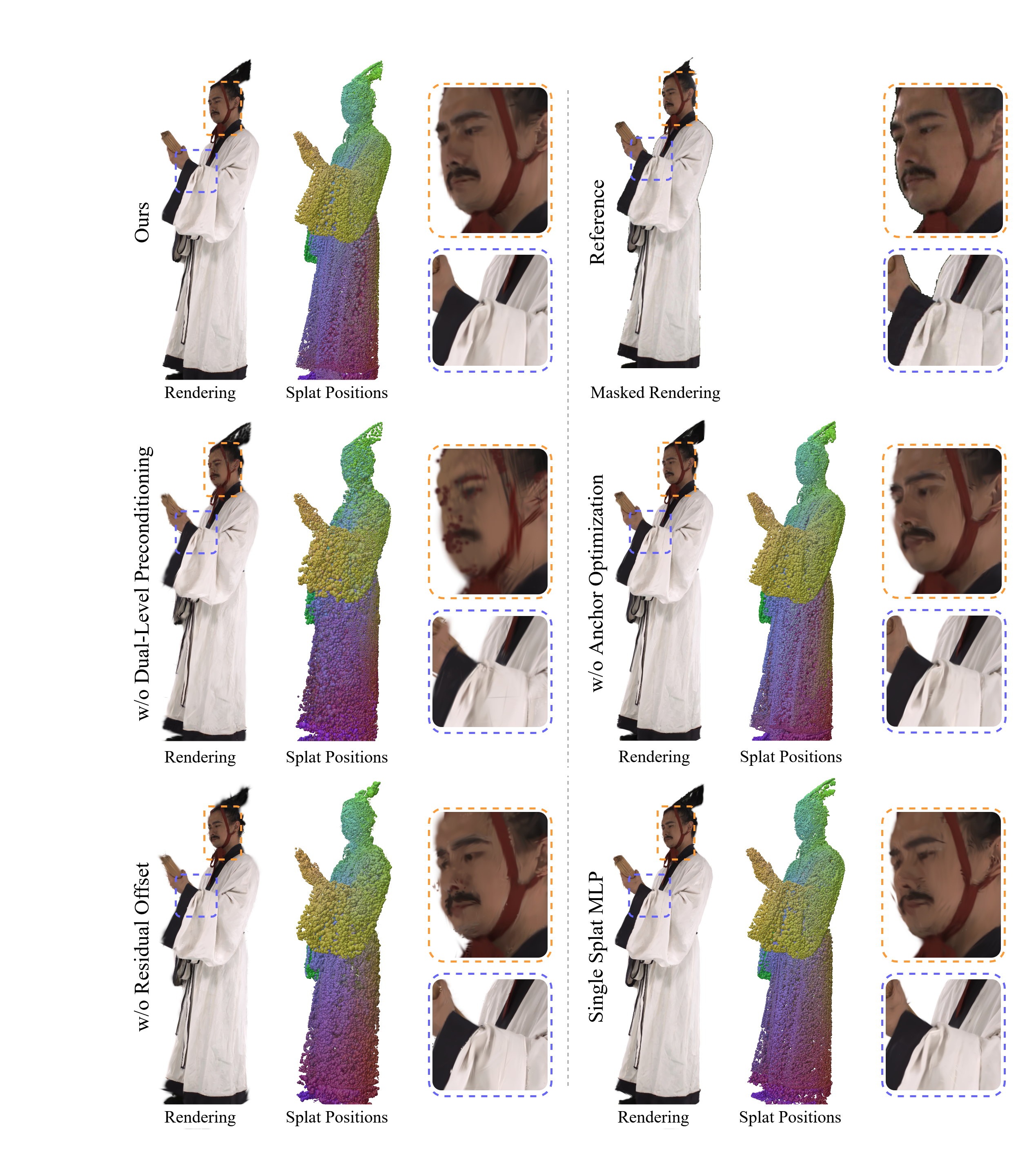
    }
    \caption{Qualitative structural ablations on DNA. Compared to our full model, removing residual offsets, anchor optimization, dual-level preconditioning, or the disentangled shape/color decoders degrades local detail and splat organization, highlighted in the face and sleeve regions.}
    \label{fig:method_ablation}
\end{figure}

\begin{table}[t]
    \centering
    \caption{Structural ablation study on DNA-dataet subject 0102, novel views. Red and orange boxes indicate best and second-best results, respectively.}
    \label{tab:ablation_structural}
    \renewcommand{\arraystretch}{1.1}
    \setlength{\tabcolsep}{3pt}
    \resizebox{0.9\linewidth}{!}{%
    \begin{tabular}{l|cccc}
        \toprule                                  & PSNR$\uparrow$ & SSIM$\uparrow$ & LPIPS$\downarrow$ & FID$\downarrow$ \\
        \midrule 
        w/o Dual-Level Preconditioning            & 28.17                     &\colorbox{second}{0.977}                       &0.054                          &24.54                        \\
        w/o Anchor Optimization                   & \colorbox{second}{28.45}                     &\colorbox{best}{0.981}                       &\colorbox{second}{0.047}                          & \colorbox{second}{13.50}                        \\
        w/o Residual Position Offsets                         &  27.24                     &0.975                       &0.053                          &36.51                       \\
        Single Splat MLP                          & 28.29                    & \colorbox{second}{0.977}                       &0.054                          &24.54                        \\
        \midrule Ours                             & \colorbox{best}{28.61}        & \colorbox{best}{0.981}        & \colorbox{best}{0.046}           & \colorbox{best}{12.66}                  \\
        \bottomrule
    \end{tabular}}
\end{table}

\paragraph{Robustness to Noisy SMPL-X Parameters}
To evaluate the robustness of our model-based formulation, we analyze how well the representation compensates for inaccuracies in the parametric body model, which in practical settings are often estimated by external pose reconstruction methods and may contain non-negligible errors. We simulate imperfect model estimates by perturbing the SMPL-X pose and shape parameters with additive zero-mean Gaussian noise of increasing magnitude and use the perturbed parameters as input to the model decoder, while following the same training and evaluation protocol as in the main experiments.

\begin{figure}[t]
    \centering
    \includegraphics[width=\linewidth, , trim=10 50 20 20, clip]{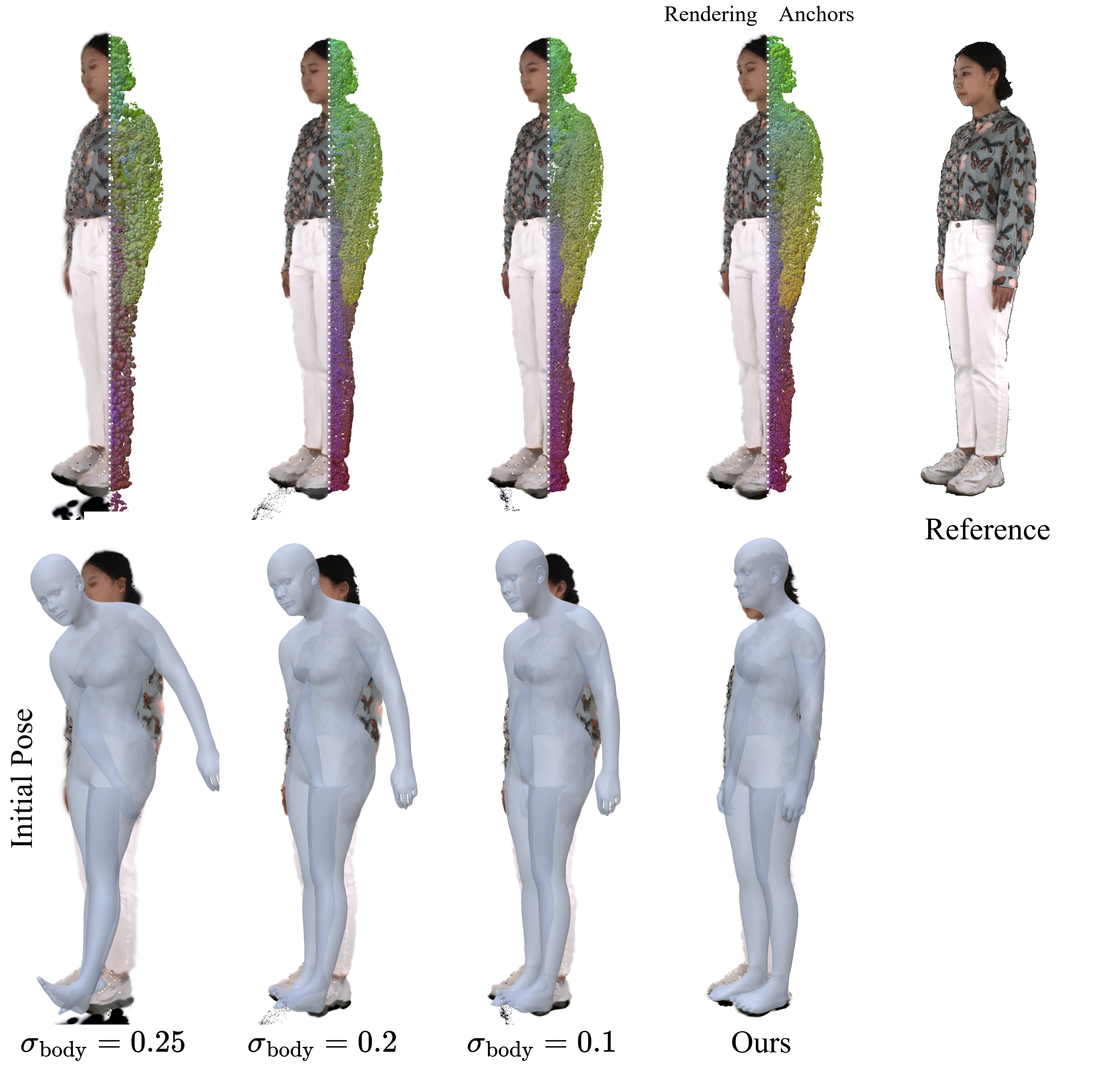}
    \caption{ Robustness to noisy SMPL-X parameters. Left: ground-truth image.
    Middle: noisy SMPL-X mesh overlay. Right: reconstruction produced by our method
    using the same noisy input. Despite significant perturbation of pose and
    shape parameters, our formulation corrects misalignments and recovers accurate
    geometry.}
    \label{fig:noise_ablation}
\end{figure}

Quantitative results in \Cref{tab:noise} show that performance degrades gracefully as the noise level increases, indicating that the learnable canonical anchors and residual splat offsets effectively compensate for moderate model inaccuracies. The canonical refinement corrects systematic shape deviations, and the residual dynamic offsets mitigate pose-dependent misalignments. Qualitative results in \Cref{fig:noise_ablation} further illustrate that even under noticeable perturbations of the SMPL-X parameters, PiG-Avatar recovers accurate surface alignment and appearance, demonstrating strong robustness to realistic model estimation errors.

\subsection{Limitations}

For fine-grained articulated parts of the human body model, particularly the fingers in SMPL-X, the actual pose may only be partially recovered from the input estimate, as local learned deformations can also capture the apparent articulation. These structures are on a similar geometric scale as the local deformations, which makes full disentanglement challenging.

In addition, PiG-Avatar relies on persistent anchor-to-surface correspondences to transport canonical splats through the articulated model. While this generally ensures stable motion under typical deformations, it can be challenged when anchors become ambiguous due to sustained interactions between topdistant body parts. For example, if a subject holds their hands at their hips throughout the sequence, anchors on the hands and hips can become less distinguishable, potentially causing minor misalignment or artifacts.

Furthermore, because we do not perform anchor densification, the anchor resolution at some regions may be lower than ideal for representing very fine details.

\begin{table}[t]
    \centering
    \caption{Robustness to noisy SMPL-X parameters on DNA-Rendering subject 0012, evaluated on novel views. Increasing $\sigma_{\mathrm{body}}$ indicates stronger body pose perturbations. PiG-Avatar degrades gracefully under moderate noise, demonstrating robustness to imperfect parametric model initialization.}
    \label{tab:noise}
    \renewcommand{\arraystretch}{1.1}
    \setlength{\tabcolsep}{3pt}
    \resizebox{0.6\linewidth}{!}{%
    \begin{tabular}{c|ccccc}
\toprule
$\sigma_{\mathrm{body}}$  & PSNR$\uparrow$ & SSIM$\uparrow$ & LPIPS$\downarrow$ & FID$\downarrow$ \\
\midrule
\phantom{0.0}0 / \phantom{00.}0$^\circ$             & 29.33 &0.977  &0.030  &18.07  \\
\phantom{0}0.1 / \phantom{0}5.7$^\circ$         &  29.12&0.977  &0.030  & 17.00  \\
\phantom{0}0.2 / 11.3$^\circ$         & 28.67 &0.976  &0.031  &17.84  \\
0.25 / 14.3$^\circ$        & 28.67 &0.974  &0.034  &27.57  \\
\bottomrule
\end{tabular}
}
\end{table}

\section{Conclusion}

We introduced PiG-Avatar, a method for reconstructing high-fidelity, animatable Gaussian avatars from multi-view video. By representing geometry in a canonical neural field and decoupling the avatar representation from the parametric body model, which is used solely for kinematic guidance, our approach avoids the geometric constraints of template-based methods. Canonical anchors sample the neural field and adapt their spatial distribution while maintaining temporal coherence, enabling stable optimization and dense, consistent surface correspondences. This design allows PiG-Avatar to accurately capture complex clothing, layered surfaces, and non-rigid motion while remaining robust to imperfect body model initialization. Our experiments demonstrate competitive perceptual quality, real-time rendering at multiple levels of detail, and reliable generalization to challenging poses and novel views, highlighting the versatility and effectiveness of our hierarchical canonical Gaussian representation.


\section*{Acknowledgements}
This work was supported by the European Regional Development Fund (ERDF) and the State of North Rhine-Westphalia as part of the operational program EFRE/JTF-Programm NRW 2021-2027. The project, titled ``Gen-AIvatar'', was funded under the NEXT.IN.NRW competition with the grant agreement No. EFRE-20801085.

Additionally, it has been funded by the Federal Ministry of Education and Research of Germany and the state of North-Rhine Westphalia as part of the Lamarr-Institute for Machine Learning and Artificial Intelligence and by the Federal Ministry of Education and Research under Grant No. 01IS22094A WEST-AI.

The work has also been funded by the Ministry of Culture and Science North Rhine-Westphalia under grant number PB22-063A (InVirtuo 4.0: Experimental Research in Virtual Environments), and by the state of North Rhine Westphalia as part of the Excellency Start-up Center.NRW
(U-BO-GROW) under grant number 03ESCNW18B.

\bibliographystyle{ACM-Reference-Format}
\bibliography{main.bib}
\end{document}